\documentclass[%
groupedaddress,
nofootinbib,
 amsmath,amssymb,
 aps, %physrev,
 prl,
 reprint,
]{revtex4-2}

\usepackage{color}             %for \textcolor
\usepackage{csquotes}        %enquote
\usepackage{soul}
\usepackage{booktabs}

\definecolor{tab10Green}{rgb}{0.1726,0.6275,0.1726}
\definecolor{tab10Red}{rgb}{0.8392,0.1530,0.1569}
\definecolor{tab10Blue}{rgb}{0.1216,0.4667,0.7059}
\definecolor{tab10Orange}{RGB}{255,127,14}
\definecolor{tab10Gray}{RGB}{127,127,127}
\definecolor{brightViolet}{RGB}{180,70,255}
\definecolor{darkgreen}{RGB}{0,127,0}

\usepackage{graphicx}% Include figure files
\usepackage{dcolumn}% Align table columns on decimal point
\usepackage{bm}% bold math
\usepackage{comment}

\usepackage[%
    pdfusetitle,
    colorlinks=true,
    citecolor=tab10Green,
    linkcolor=tab10Red,
    urlcolor=tab10Blue,
]{hyperref}% add hypertext capabilities
\usepackage{braket}            %for braket notation

\newcommand{\Fig}[1]{Figure \ref{#1}}

\newcommand{\Refl}[1]{Ref.~\cite{#1}}    %Might need to make more often than usual
\DeclareRobustCommand{\eqnr}[1]{Eq.~\eqref{#1}}
\begin{document}

\title{On the effective restoration of $U(1)_A$ symmetry at finite temperature 
}% 

\author{Gert Aarts, Chris Allton}
\affiliation{Centre for Quantum Fields and Gravity, Department of Physics, Swansea University, Swansea, SA2 8PP, United Kingdom}
 \author{Ryan Bignell}
 \email{bignellr@tcd.ie}
 \affiliation{School of Mathematics \& Hamilton Mathematics Institute, Trinity College Dublin, Dublin, Ireland}
\author{Benjamin J\"ager}
\affiliation{%
 Quantum Field Theory Center \& Danish IAS, Department of Mathematics and Computer Science \\
  University of Southern Denmark, 5230, Odense M, Denmark
}%
\author{Seyong Kim}
\affiliation{
Department of Physics, Sejong University, Seoul 05006, Korea
}%
\author{Jon-Ivar Skullerud}
\affiliation{
 Department of Physics, National University of Ireland Maynooth, County Kildare, Ireland
}
\affiliation{School of Mathematics \& Hamilton Mathematics Institute, Trinity College Dublin, Dublin, Ireland}
\author{Antonio Smecca}
\email{antonio.smecca@roma3.infn.it}
\affiliation{%
 INFN,
  Sezione di Roma Tre, Via della Vasca Navale 84, I-00146 Rome, Italy
}%
\affiliation{Centre for Quantum Fields and Gravity, Department of Physics, Swansea University, Swansea, SA2 8PP, United Kingdom}

\date{\today}

\begin{abstract}
The $U(1)_A$ symmetry of the massless QCD Lagrangian is explicitly broken by the axial anomaly, but it may be effectively restored at finite temperature. Determining the temperature at which this occurs is important for understanding the chiral transition and the structure of the QCD phase diagram. A commonly used probe of effective $U(1)_A$ restoration is the degeneracy of flavour non-singlet pseudoscalar and scalar susceptibilities. Using anisotropic lattice QCD ensembles with Wilson-clover fermions generated by the \textsc{Fastsum} collaboration, we study this degeneracy through hadronic correlation functions over a wide range of temperatures. The fine temporal resolution of our Generation 3 ensembles allows us to determine the temperature at which the pseudoscalar and scalar channels become degenerate. We find evidence for the effective restoration of $U(1)_A$ symmetry at $T_{U(1)_A}=319(22)$ MeV, well above the chiral crossover temperature.
\end{abstract}

%\keywords{Suggested keywords}%Use showkeys class option if keyword display desired

\maketitle

\paragraph{Introduction --} 
During the last fifty years, it has been established that quantum chromodynamics (QCD) is the correct theory of the strong interaction~\cite{Gross:2022hyw}. Despite this, the behaviour of QCD under the extreme conditions of high temperature and density remains only partially understood \cite{Aarts:2023vsf}. Of particular interest is the study of thermal phase transitions, which in certain limits are associated with the restoration of symmetries of the QCD Lagrangian \cite{Aarts:2023vsf,Borsanyi:2025ttb}.
The massless QCD Lagrangian is invariant under the symmetry $U(1)_V\times U(1)_A\times SU(n_f)_L\times SU(n_f)_R$, where $n_f$ denotes the number of quark flavours. In vacuum, $SU(n_f)_L \times SU(n_f)_R$ is spontaneously broken, giving rise to massless Goldstone modes. Below we refer to this as chiral symmetry breaking.
On the other hand, the $U(1)_A$ symmetry is explicitly broken in the quantised theory~\cite{Bell1969,Adler1} and is responsible for the higher mass of the $\eta'$ meson through the Witten-Veneziano formula~\cite{Witten:1979vv,VENEZIANO1979213}. In nature, light quarks have small but non-zero masses, making the above symmetries approximate.

It is now understood that at high temperature chiral symmetry is restored~\cite{Aoki:2006br,HotQCD:2019xnw,Cuteri:2021ikv}. For physical quark masses the transition is a crossover occurring at $T_{c}\approx154$ $\mathrm{MeV}$~\cite{HotQCD:2018pds,Borsanyi:2020fev,Gavai:2024mcj}. 
Restoration of chiral symmetry implies the degeneracy of hadronic channels related through a chiral rotation. 
Well-known examples of such chiral partners are the vector ($\rho$) and axial-vector ($a_1$) mesons, the flavour singlet scalar ($\sigma$, also known as $f_0$) and the flavour non-singlet pseudoscalar ($\pi$) mesons, and, in the nucleon sector, the $N(939)$ and $N^*(1535)$ states.
For light and strange baryons, the emergence of parity doubling, which indicates chiral symmetry restoration, has indeed been observed using lattice QCD \cite{Aarts:2015mma,Aarts:2017rrl,Aarts:2018glk}.

Although $U(1)_A$ symmetry is broken explicitly by the anomaly, it is expected to be effectively restored at high temperature, which can also be studied at the level of correlation functions \cite{Shuryak:1993ee,Cohen:1996ng,Cohen:1997hz}. 
In this case, the restoration would imply a degeneracy between the flavour non-singlet pseudoscalar ($\pi$) and scalar ($\delta$, also known as $a_0$) channels. 
Note that by effective restoration we mean that 
differences between related channels due to $U(1)_A$ symmetry breaking become negligible.
The realisation of the effective restoration is still an open question with important consequences for the order of the chiral transition in the chiral limit and the structure of the QCD phase diagram~\cite{PhysRevD.29.338,Brandt:2016daq,PhysRevD.91.094504}. $U(1)_A$ restoration is an intricate part of QCD, as it relates the Dirac operator spectral density~\cite{Aoki:2012yj,PhysRevD.91.094504} and the contribution of near-zero modes above $T_c$ with the topology of the gauge fields \cite{Bonanno:2025wcv}.

Due to the nonperturbative nature of QCD, effective $U(1)_A$ symmetry restoration at nonzero temperature has been studied using lattice simulations, comparing either susceptibilities in the flavour non-singlet pseudoscalar and scalar channels, or by analysing near-zero modes of the Dirac operator; for recent reviews, see Refs.~\cite{Lahiri:2021lrk,Ding:2026gao}. Given the importance of chiral symmetries, most investigations have used a lattice fermion formulation with unbroken chiral symmetry at non-zero lattice spacing \cite{Narayanan:1993sk,Narayanan:1994gw,Brower:2012vk}. 
The present picture remains ambiguous \cite{Gross:2022hyw,Ding:2026gao}, despite the longstanding effort of several groups. Some results suggest that $U(1)_A$ is effectively restored near $T_c$ \cite{Cossu:2013uua,Tomiya:2016jwr,PhysRevD.88.019901}, while others do not find this \cite{Buchoff:2013nra,PhysRevD.91.094504,Alexandru:2024tel,Kaczmarek:2023bxb,HotQCD:2012vvd,Kaczmarek:2021ser,Gavai:2024mcj,Bhattacharya:2014ara,Ding:2020xlj}. This matter is extensively discussed in Refs.~\cite{Giordano:2025shr,Giordano:2025fcr,Giordano:2025vbb}, clarifying assumptions made in earlier work.
Another element to consider is the role of the strange quark~\cite{Cianti:2025uzz,GomezNicola:2020qxo,Borsanyi:2025ttb}.

Wilson lattice fermions break chiral symmetry explicitly. Hence their use to investigate the restoration of $U(1)_A$ symmetry has been limited. An exception is Ref.~\cite{Brandt:2016daq}, in which screening masses in the relevant channels were studied. 
Without explicit chiral symmetry, two aspects should be considered \cite{Capitani:2002mp}.
First, at finite lattice spacing the effect of the Wilson term is prominent at small distances. Second, the renormalisation of operators is not protected by chiral symmetry. 
In this work, we present a study using Wilson-clover fermions~\cite{Wilson:1974sk,Sheikholeslami:1985ij}, on ensembles with an unprecedented fine lattice spacing in the temporal direction. We address these two aspects in detail.
By analysing temporal correlation functions in the flavour non-singlet pseudoscalar and scalar channels, our main finding is that $U(1)_A$ symmetry is effectively restored at $T\gtrsim 300\,$MeV. 

\begin{table}[]
\caption{Details of the \textsc{Fastsum} Generation 2, 2L and 3 ensembles with $n_f=2+1$ flavours on $N_s^3\times N_\tau$ lattices: the number of lattice points $N_s$ in the spatial direction ($^*$in Generation 3, $N_s=24$ for $N_\tau=256$); the lattice spacings $a_s, a_\tau$; the renormalised anisotropy $\xi = a_s/a_\tau$; the pion mass $m_\pi$; the chiral transition temperature $T_c$ determined using the renormalised chiral condensate. The strange quark mass is approximately at its physical value.
}
\label{tab:lattice}
\begin{tabular}{l | c | c | c}
\toprule
                    & Generation 2  & Generation 2L & Generation 3 \\
                    \midrule
$N_s$               & 24        & 32         & 32 ($^*$24) \\
$a_s$ (fm)          & 0.1205(8) & 0.1121(3)  & 0.1080(6) \\
$a_\tau^{-1}$ (GeV) & 5.63(4)   & 6.079(13)  & 12.82(9) \\
$\xi = a_s/a_\tau$  & 3.444(6)  & 3.453(6)   & 7.02(3)   \\ 
$m_\pi$ (MeV)       & 384(4)    & 239(1)     & 380(1)  \\
$T_c$ (MeV)         & 181(1)    & 167(3)     & 182(1)  \\
Refs.               & \cite{Aarts:2014nba,aarts_2023_8403827} & \cite{Aarts:2020vyb,aarts_2024_10636046} & \cite{Skullerud:2025xva,Gen3} \\
\bottomrule
\end{tabular}

\caption{Temporal extent $N_\tau$ of the lattices in the three generations, with the corresponding temperature $T$ expressed in MeV and in units of $T_c$.
}
\label{tab:ensembles}
\begin{tabular}{ccc|ccc|ccc}
\toprule
\multicolumn{3}{c|}{Generation 2} & \multicolumn{3}{c|}{Generation 2L} & \multicolumn{3}{c}{Generation 3} \\ \midrule
$N_\tau$ & $T$ (MeV) & $T / T_{c}$ & $N_\tau$ & $T$ (MeV) & $T / T_{c}$ & $N_\tau$ & $T$ (MeV) & $T / T_{c}$ \\ \midrule
128 & 44 & 0.24 & 128 & 47 & 0.42 & 256 & 50 & 0.28 \\
 &  &  & 64 & 95 & 0.59 & 128 & 100 & 0.55 \\
 &  &  & 56 & 109 & 0.67 & 112 & 114 & 0.64 \\
48 & 117 & 0.63 & 48 & 127 & 0.78 & 96 & 133 & 0.73 \\
40 & 141 & 0.76 & 40 & 152 & 0.94 & 80 & 160 & 0.88 \\
\multicolumn{1}{l}{} & \multicolumn{1}{l}{} & \multicolumn{1}{l|}{} & \multicolumn{1}{l}{} & \multicolumn{1}{l}{} & \multicolumn{1}{l|}{} & 76 & 169 & 0.93 \\
36 & 156 & 0.84 & 36 & 169 & 1.04 & 72 & 178 & 0.98 \\
\multicolumn{1}{l}{} & \multicolumn{1}{l}{} & \multicolumn{1}{l|}{} & \multicolumn{1}{l}{} & \multicolumn{1}{l}{} & \multicolumn{1}{l|}{} & 68 & 188 & 1.04 \\
32 & 176 & 0.95 & 32 & 190 & 1.17 & 64 & 200 & 1.10 \\
28 & 201 & 1.09 & 28 & 217 & 1.34 & 56 & 229 & 1.26 \\
24 & 235 & 1.27 & 24 & 253 & 1.56 & 48 & 267 & 1.47 \\
20 & 281 & 1.52 & 20 & 304 & 1.87 & 40 & 320 & 1.76 \\
\multicolumn{1}{l}{} & \multicolumn{1}{l}{} & \multicolumn{1}{l|}{} & \multicolumn{1}{l}{} & \multicolumn{1}{l}{} & \multicolumn{1}{l|}{} & 36 & 356 & 1.96 \\
16 & 352 & 1.90 & 16 & 380 & 2.34 & 32 & 400 & 2.20 \\
 &  &  & 12 & 507 & 3.12 & 24 & 534 & 2.93 \\
 &  &  & 8 & 760 & 4.55 &  &  & \\
 \bottomrule
\end{tabular}
\end{table}

\paragraph{Lattice ensembles --}
The anisotropic thermal ensembles of the \textsc{Fastsum} collaboration span a wide range of temperatures above and below the chiral transition temperature $T_c$. We use a fixed-scale approach, in which the temperature $T$ is varied by changing the number of points $N_\tau$ in the Euclidean time direction, $T=1/(a_\tau N_\tau)$. The anisotropy is the ratio of the lattice spacing in the space and time directions, $\xi = a_s/a_\tau$, and is larger than 1.
Our lattice action matches that of the Hadron Spectrum Collaboration~\cite{Edwards:2008ja} and is a Symanzik improved anisotropic gauge action with tree-level mean-field coefficients and $n_f=2+1$ flavours of mean-field-improved Wilson-clover with stout-smeared links~\cite{Morningstar:2003gk}.

Simulations directly at the physical point are not yet feasible. Instead we follow a strategy in which we have generated ensembles in which the degenerate light quark masses and the temporal lattice spacing are varied, whereas the strange quark is tuned to approximately its physical value 
\cite{HadronSpectrum:2008xlg,HadronSpectrum:2012gic,Cheung:2016bym}.
In detail, we generated three sets of ensembles: Generations 2 and 3 have approximately the same pion mass, $m_\pi\approx380$ MeV, but differ by a factor two in anisotropy, $\xi\approx 3.4$ and $7.0$ respectively. Generations 2 and 2L have almost the same anisotropy but differ in pion mass, $m_\pi=384(4)$ and $239(1)$ MeV respectively.
An overview is given in Tables \ref{tab:lattice} and \ref{tab:ensembles}. 
 Note that we use the updated lattice spacing of \Refl{Wilson:2019wfr} for Generation 2L. 
The chiral transition temperature depends on the pion mass; for a comparison between Generation 2 and 2L, see Ref.~\cite{Aarts:2020vyb}.
Generation 3 is presented here for the first time (preliminary studies were shown in Refs.~\cite{Skullerud:2025xva,Bignell:2026ybw}) and will be detailed further in future work~\cite{Gen3}. As can be seen in Table \ref{tab:ensembles}, Generation 3 has an unparalleled temperature coverage. A comparison between Generation 2 and 3 also permits an understanding of the approach to the temporal continuum limit.

\paragraph{Correlators and susceptibilities --}
We start with the (connected) Euclidean correlation functions of meson operators at temperature $T$, projected to zero momentum,
\begin{align}
    G(\tau, T) = \sum_{\boldsymbol{x}} \braket{ \mathcal{O}(\tau,\boldsymbol{x})\,\mathcal{O}^\dagger(0,\boldsymbol{0})}_c,
    \label{eq:sus-default}
\end{align}
where $\mathcal{O}=\overline{\psi}\Gamma\psi$ with $\Gamma = \left(\gamma_5,\,\mathcal{I}\right)$ for the $\left(\pi,\,\delta\right)$-mesons respectively.
Including the disconnected contributions in the scalar channel would yield the flavour singlet $\sigma$ (or $f_0$) meson. As stated, here we consider the flavour non-singlet $\delta$ (or $a_0$) meson.
Several studies \cite{HotQCD:2012vvd,Buchoff:2013nra,RevModPhys.65.1,Cossu:2013uua,Suzuki:2019vzy,Brandt:2016daq,Bhattacharya:2014ara,Bazavov:2019www} have investigated the restoration of $U(1)_A$ symmetry by examining the susceptibilities
\begin{equation}
    \begin{aligned}
\chi_{(\pi,\delta)}(T) &= \frac{1}{2}\int d^4x\,\braket{\mathcal{O}_{(\pi,\delta)}(x)\mathcal{O}^{\dagger}_{(\pi,\delta)}(0)}_c \\
&= \frac{1}{2} \sum_{\tau}G_{(\pi,\delta)}(\tau, T).
\label{eq:sus}
\end{aligned}
\end{equation}
As flavour quantum numbers are invariant under $U(1)_A$ transformations, the degeneracy of these two susceptibilities indicates the effective restoration of $U(1)_A$ symmetry~\cite{HotQCD:2012vvd} and the difference $\chi_{\pi} - \chi_{\delta}$ can be used to study this.
Susceptibilities are a good probe of $U(1)_A$ symmetry restoration when a good chiral action is used in the simulation. However, even chirally symmetric actions such as domain wall fermions suffer from lattice artefacts affecting the correlators at short distances~\cite{Cossu:2013uua,HotQCD:2012vvd,Suzuki:2019vzy}. 

In our setup, the Wilson term explicitly breaks chiral symmetry. This predominantly affects the short-distance part of the correlator~\cite{Capitani:2002mp}. The susceptibility difference is expected to suffer from contact term contributions~\cite{Brandt:2016daq}. Moreover, operator renormalisation is not protected by chiral symmetry. To mitigate these effects, we first consider the ratio
\begin{align}
\widetilde{G}_{(\pi,\delta)}(\tau,T)=\frac{G_{(\pi,\delta)}(\tau,T)}{G_{(\pi,\delta)}(N_\tau/2,T)},
\end{align}
in which the renormalisation factor cancels (here and below, integer $\tau$ is Euclidean time in units of $a_\tau$). 
Subsequently we restrict the sum over Euclidean time and define the (dimensionless) normalised susceptibilities
\begin{align}
\widetilde{\chi}_{\left(\pi,\,\delta\right)}^{\mathrm{norm}}(T,\tau_{\min}) = \frac{1}{N_{\tau}} \sum_{\tau=\tau_{\min}}^{N_{\tau}/2} \widetilde G_{\left(\pi,\,\delta\right)}(\tau,T).
\label{eq:norm_sus}
\end{align}
Here we used that $G(N_\tau-\tau)=G(\tau)$ for meson channels.
The temporal cutoff $\tau_{\text{min}}$ removes the unwanted short distance effects coming from the Wilson term. In addition, we consider both \enquote{local} and \enquote{smeared} correlators, where \enquote{smeared} correlators are expected to have reduced overlap with excited states and with artefacts from the Wilson mass term; see \hyperref[sec:endmatter]{End Matter} for details. The conjecture is that these normalised susceptibilities are infrared dominated observables, sensitive to the restoration of the $U(1)_A$ symmetry. The choice of $\tau_{\min}$ depends on where the low-energy dominance sets in. Varying the value of $\tau_{\text{min}}$ gives an indication of the systematic uncertainty in the analysis.

A slightly alternative way to implement the same idea is to consider 
\begin{align}
    R_{\pi-\delta}(\tau, T) = \frac{\widetilde{G}_{\pi}(\tau, T) - \widetilde{G}_{\delta}(\tau, T)}{\widetilde{G}_{\pi}(\tau, T) + \widetilde{G}_{\delta}(\tau, T)}.
  \label{eq:R_ratio}
\end{align}
By construction, the value of $R_{\pi-\delta}(\tau, T)$ is zero at $\tau=N_{\tau}/2$, while for other values of $\tau$ this ratio only vanishes when the two normalised correlators are degenerate.

\begin{figure}[t]
   \centering
    \includegraphics[width=1.\linewidth]{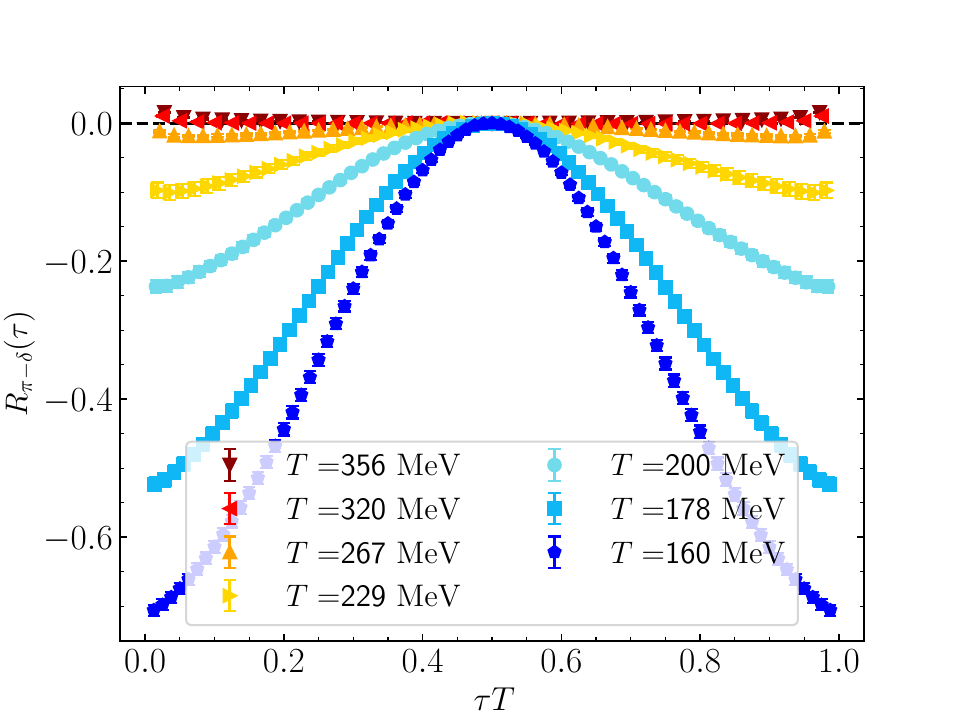}
    \caption{Normalised ratio $R_{\pi-\delta}(\tau, T)$ of the difference between normalised $\pi$ and $\delta$ correlators, see Eq.~(\ref{eq:R_ratio}), for the Generation 3 ensembles using smeared correlators.}
    \label{fig:PS_R_ratio_smeared}
\end{figure}

The results for $R_{\pi-\delta}(\tau, T)$ are shown in Figure~\ref{fig:PS_R_ratio_smeared} for the Generation 3 ensembles. These results (and all those discussed below) are obtained with smeared operators at source and sink; a detailed analysis of smearing is presented in the \hyperref[sec:endmatter]{End Matter}. At low temperature we observe an inverted parabola near the centre of the lattice. This can be understood as follows. 
If we assume a single-pole Ansatz for the meson correlators, with dimensionless masses $\tilde{m}_{(\pi,\delta)}=m_{(\pi,\delta)}/T$, then \eqnr{eq:R_ratio} becomes
\begin{align}
\begin{aligned}
        R_{\pi-\delta}(\tau, T) &= \frac{\cosh{\tilde{m}_{\pi}\,x} - \cosh{\tilde{m}_{\delta}\,x}}{\cosh{\tilde{m}_{\pi}\,x} + \cosh{\tilde{m}_{\delta}\,x}}  \\
    &= \frac{1}{4}\,\left(\tilde{m}_{\pi}^2 - \tilde{m}^2_{\delta}\right)\,x^2 + \mathcal{O}\left(x^4\right),
    \end{aligned}
\end{align}
where $x = \tau\,T - 1/2$ and the expansion is valid for small $x$, i.e., near the centre of the lattice. 
Since $m_\delta>m_\pi$, the curvature is negative. 
As the temperature increases, we observe that the curvature decreases and that the ratio becomes consistent with zero at the two highest temperatures, indicating a degeneracy between the flavour non-singlet pseudoscalar and scalar channels.

\begin{figure}[t]
    \centering
    \includegraphics[width=1.\linewidth]{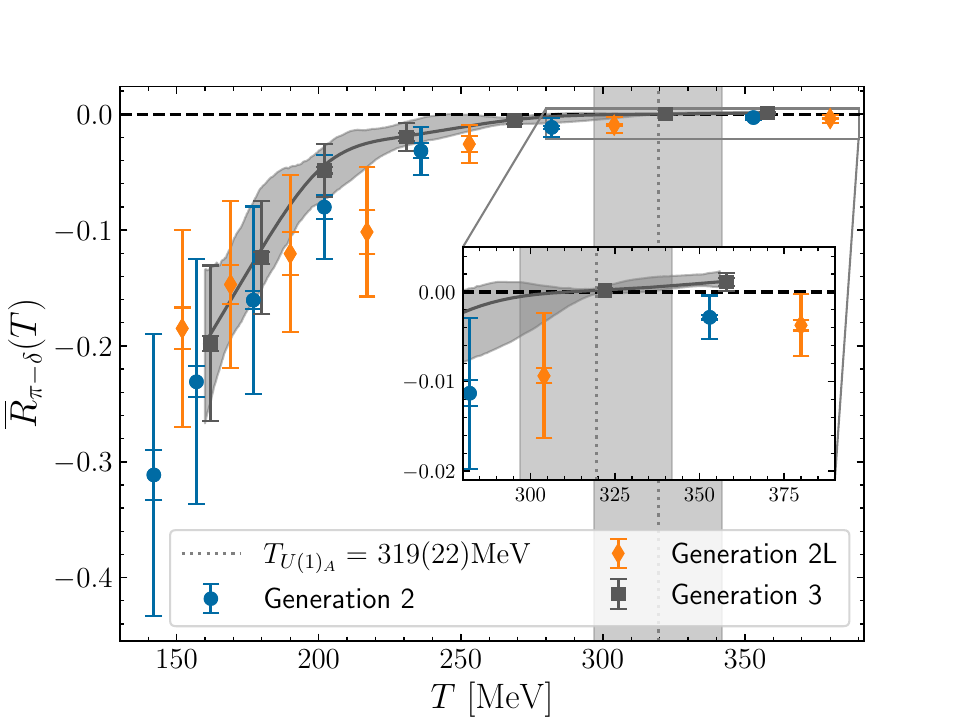}
    \caption{$\overline{R}_{\pi-\delta}$ results for Generation 2 $(\text{blue circles }\textcolor{tab10Blue}{\bigcirc})$, 2L $(\text{orange diamonds }\textcolor{tab10Orange}{\diamond})$ and 3 $(\text{dark-gray squares }\textcolor{tab10Gray}{\square})$, with the inset showing a close-up of the region of interest. The dark-gray curve shows a cubic spline interpolation of Generation 3 data only and $T_{U(1)_A}$ is taken as the temperature where the spline reaches zero within errors.}
    \label{fig:delta_chi}
\end{figure}

To probe the $U(1)_A$ restoration in detail, we consider the normalised difference of the normalised susceptibilities (\ref{eq:norm_sus}),
\begin{align}
    \overline{R}_{\pi-\delta}(T, \tau_{\min}) = \frac{ \widetilde{\chi}^{\mathrm{norm}}_{\pi}(T, \tau_{\min}) - \widetilde{\chi}^{\mathrm{norm}}_{\delta}(T, \tau_{\min})}{\widetilde{\chi}^{\mathrm{norm}}_{\pi}(T, \tau_{\min}) + \widetilde{\chi}^{\mathrm{norm}}_{\delta}(T, \tau_{\min})}.
\end{align}
Using Eq.~(\ref{eq:R_ratio}) summed over $\tau_{\rm min}\leq \tau\leq N_\tau/2$ gives very similar results. 
Similar ratios were considered in Refs.~\cite{Datta:2012fz,Aarts:2015mma,Aarts:2017rrl,Aarts:2018glk,Smecca:2024gpu,Chiu:2026sxy}. Again, this quantity vanishes when $U(1)_A$ is effectively restored.
The dependence of $\overline{R}_{\pi-\delta}(T, \tau_{\min})$ on the temperature is shown in Figure~\ref{fig:delta_chi}, for all three generations. We observe that the ratio approaches zero far above the chiral transition temperature. This is our first, qualitative, finding.

To obtain a quantitative result, including an estimate of the systematic uncertainties, we proceed as follows.  
To estimate the uncertainty due to the choice of $\tau_{\min}$, we compute $\overline{R}_{\pi-\delta}$ for several values of $\tau_{\min}$ within the interval $\tau_{\min}T\simeq[0.125,0.44]$ for Generation 2 and 2L and $\tau_{\min}T\simeq[0.13,0.36]$ for Generation 3. 
The central value is taken as the midpoint between the maximum and minimum values obtained in these intervals, and the corresponding systematic uncertainty is defined as
\[
   \sigma_{\mathrm{sys}}(T)=\frac{1}{2} \left( \overline{R}_{\pi-\delta}(T,\tau_{\min})\big|_{\min}-\overline{R}_{\pi-\delta}(T,\tau_{\min})\big|_{\max} \right).
\] 
The total error is obtained adding the statistical and systematic errors in quadrature.
The smaller caps on the error whiskers in Figure~\ref{fig:delta_chi} indicate the statistical error and the larger caps the total error.
Our conclusions are robust under variations of the temporal cutoff $\tau_{\min}$ and across the three generations.

Since the $U(1)_A$ symmetry is only expected to be effectively restored, with no phase transition, any criterion for determining the restoration temperature will necessarily be somewhat arbitrary. One option is to define it as where $\overline{R}_{\pi-\delta}$ decreases to below a certain fraction of its zero-temperature value, prompting the question of which fractional value to choose, with results depending on this choice.  Here we have opted instead for the stricter condition that $\overline{R}_{\pi-\delta}=0$ within errors at the restoration temperature. 
Selecting an alternative criterion, in which $\overline{R}_{\pi-\delta}=0$ to within a certain fraction, does not affect the main conclusion, namely that $U(1)_A$ symmetry is effectively restored considerably above the chiral transition.
To determine the restoration temperature, we interpolate the central values of $\overline{R}_{\pi-\delta}$ at different temperatures with a cubic spline, and use this to find the zero crossing point.
The uncertainty on $T_{U(1)_A}$ is estimated via a Monte Carlo resampling procedure, in which the input data are sampled according to their uncertainties and the spline interpolation is carried out for each replica.

We find that the Generation 3 results indeed reach zero within the quoted uncertainty. On the other hand, the Generation 2 and 2L results get close to zero, but they do not exhibit a clear crossing, and central values lie within approximately one standard deviation from zero. We take this as being due to the smaller temporal lattice spacing of Generation 3, with twice the anisotropy compared to Generation 2 and 2L. With this higher resolution in the temporal direction, the Generation 3 data provide clear evidence for effective restoration of $U(1)_A$ symmetry at $T_{U(1)_A}=319(22)$ $\mathrm{MeV}$. This is our second, quantitative, finding. From our results we cannot conclude whether the transition temperature depends on the quark mass. 

We have carried out a similar analysis for the restoration of $SU(2)_L\times SU(2)_R$ chiral symmetry.
We find a distinctively lower temperature for the chiral transition, see the \hyperref[sec:supp]{Supplementary Material} for details. 

\paragraph{Discussion --}
Using the state-of-the-art set of ensembles produced by the \textsc{Fastsum} collaboration, we find evidence for the effective restoration of $U(1)_A$ symmetry at $T=319(22)$ $\mathrm{MeV}$. This is the first large-scale study of this problem using Wilson fermions on highly anisotropic lattices. Our results suggest the following pattern of symmetry restoration in QCD with non-zero quark masses: a crossover transition at which $SU(2)_L\times SU(2)_R$ symmetry is restored, which takes place at $T_c \approx 154$ $\mathrm{MeV}$ for physical quark masses, and the effective restoration of $U(1)_A$ symmetry at a considerably higher temperature, $T_{U(1)_A}\gtrsim300\,$MeV. 

Further evidence of restoration of $U(1)_A$ symmetry around $320$ $\mathrm{MeV}$ was recently reported in Ref.~\cite{Shanker:2026mtu}, where in an extensive study the infrared eigenstates of the Dirac operator are understood in terms of random matrix theory. Those results are consistent with other explorations~\cite{Alexandru:2019gdm,Horvath:2025ypt} of the Dirac operator eigenstates, which suggest the existence of more than one transition temperature in QCD~\cite{Hanada:2025rca,Fujimoto:2025sxx}. We note that this temperature coincides with the regime where the topological susceptibility exhibits a power-law scaling~\cite{Aarts:2023vsf,Burger:2018fvb,Petreczky:2016vrs,Kotov:2021rah}, in accordance with the dilute instanton gas approximation (DIGA)~\cite{RevModPhys.53.43}. 
An intermediate phase between the two transitions has also been related to an emergent symmetry absent in the QCD Lagrangian known as chiral-spin (CS) symmetry~\cite{Glozman:2014mka}, see also Refs.~\cite{Rohrhofer:2019qwq,Rohrhofer:2019qal,Glozman:2022lda,Chiu:2023hnm,Chiu:2024bqx,Chiu:2024jyz}.
It is similarly intriguing to note that $T_{U(1)_A}$ coincides with the proposed deconfinement transition found by analysing centre vortices~\cite{Mickley:2024vkm,Mickley:2025qgk}.

While chiral symmetry restoration occurs at different temperatures for Generation 2 and 2L, due to the different pion masses (see Table~\ref{tab:lattice} and Ref.~\cite{Aarts:2020vyb}), the $U(1)_A$ analysis presented here shows no strong quark-mass dependence within our current precision, consistent with Ref.~\cite{Bhattacharya:2014ara}. In contrast, \Refl{Ding:2020xlj} finds a mild linear dependence on the light quark mass squared in the continuum limit.
In the future we plan to study the light quark mass dependence more carefully using the \textsc{Fastsum} Generation 3L and 3P ensembles, which will have the same pion mass as in Generation 2L and a physical pion mass respectively.
The Gen3L ensembles are currently in production, while production of the Gen3P ensembles is expected to begin shortly.

\vspace*{0.2cm}

\paragraph{Open Access Statement --}
For the purpose of open
access, the authors have applied a Creative Commons
Attribution (CC BY) licence to any Author Accepted
Manuscript version arising.

\paragraph{Software and Data --}
The Generation 3 ensembles were generated using \textsc{openQCD-Fastsum}~\cite{openqcd-fastsum}, a derivative of \textsc{OpenQCD-1.6}. They will be made available at some point in the future in accordance with \textsc{Fastsum}'s sharing policy. The Generation 2 ensembles are publicly available~\cite{aarts_2023_8403827} while Generation 2L is available for non-competing research upon request~\cite{aarts_2024_10636046}. Meson correlators were computed using \textsc{openQCD-hadspec}~\cite{openqcd-hadspec}.

The correlators and analysis workflow are available in \Refl{thisZenodo}. The workflow uses Snakemake~\cite{snakemake} and is inspired by the approach of the TELOS Collaboration~\cite{Bennett:2025neg}.
Additional software analysis tools include \textsc{gvar}~\cite{peter_lepage_2025_14783421}, \textsc{FJSample}~\cite{FJSample}, and \textsc{YAAC}~\cite{YAAC}.

\paragraph{Authors' Contributions --}
\begin{itemize}
    \item Bignell, Smecca: Data production and analysis, manuscript production and physics interpretation of results
    \item J\"ager: Data production
    \item Skullerud: Data production, contribution to manuscript and physics interpretation of results.
    \item Aarts, Allton, Kim: Contributions to analysis and physics interpretations of results, and draft of the manuscript.
\end{itemize}

\paragraph{Acknowledgements -- }
This work is supported by STFC grant ST/X000648/1. GA is supported by a Royal Society Leverhulme Trust Senior Research Fellowship. RB acknowledges support from a Science Foundation Ireland Frontiers for the Future Project award with grant number SFI-21/FFP-P/10186. SK is supported by the National Research Foundation of Korea through the grant, NRF-2008-000458. AS is supported by the Italian Ministry of University and Research (MUR) projects FIS 0000155. We acknowledge the EuroHPC Joint Undertaking for awarding the projects EHPC-EXT-2023E01-010 and EXT-2025E01-079 access to LUMI-C and LUMI-G, Finland. This work used the DiRAC Data Intensive service (DIaL2 \& DIaL) at the University of Leicester, managed by the University of Leicester Research Computing Service on behalf of the STFC DiRAC HPC Facility (www.dirac.ac.uk). The DiRAC service at Leicester was funded by BEIS, UKRI and STFC capital funding and STFC operations grants. This work used the DiRAC Extreme Scaling service (Tesseract) at the University of Edinburgh, managed by the Edinburgh Parallel Computing Centre on behalf of the STFC DiRAC HPC Facility (www.dirac.ac.uk). The DiRAC service at Edinburgh was funded by BEIS, UKRI and STFC capital funding and STFC operations grants. This work was performed using PRACE resources at Joliot-Curie (Irene) hosted in Rome and Hawk hosted by HLRS Stuttgart. We acknowledge the use of computing resources from the Irish Centre for High-End Computing (ICHEC). We acknowledge the support of the Supercomputing Wales project, which is part-funded by the European Regional Development Fund (ERDF) via Welsh Government.

\clearpage

\section{End Matter}
\label{sec:endmatter}

\paragraph{Local versus smeared operators --}
To control the influence of excited state contributions and lattice artefacts from the Wilson mass term, we make use of \enquote{smeared} correlation functions. These are known to have a reduced overlap with excited states~\cite{Gusken:1989qx}.
Here we compare results obtained with \enquote{local} and \enquote{smeared} correlators. In the latter, the bare (delta function) source $\eta$ is smeared~\cite{Gusken:1989qx}, i.e., $\eta^\prime = C_{\text{norm}}\left(1 + \kappa H\right)^{n}\eta$.
Here $\kappa$ and $n$ determine the degree of smearing, $H$ is the spatial hopping part of the Dirac operator and $C_{\text{norm}}$ is an appropriate normalisation. We use a smearing of $(\kappa, n) = (1.96, 6)$ giving $r_{\text{RMS}}=2.45$. The sink is similarly smeared. 

\begin{figure}[h]
    \centering
    \includegraphics[width=0.96\linewidth]{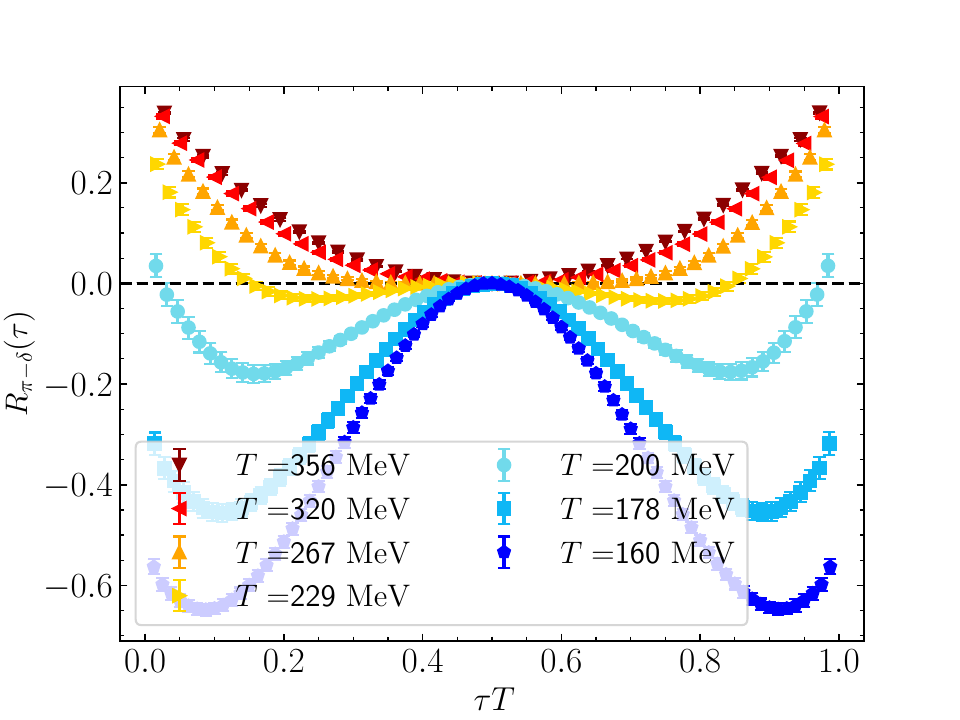}
    \caption{As \Fig{fig:PS_R_ratio_smeared}, using local sources and sinks.}
    \label{fig:PS_R_ratio_local}
\end{figure}

Our main result for the time-dependent ratio $R_{\pi-\delta}(\tau,T)$, see Eq.~(\ref{eq:R_ratio}), was shown in \Fig{fig:PS_R_ratio_smeared} for smeared sources. The equivalent using local sources is shown in Figure \ref{fig:PS_R_ratio_local}. We observe a turnover of the curvature at the centre of the lattice, but no flattening. We also observe an upward bend at the edges of the lattice
at all temperatures,
influencing the choice of $\tau_{\rm min}$.
We have determined that the lack of flattening is due to the explicit chiral symmetry breaking of the Wilson mass term, by comparing free Wilson and chirally symmetric overlap quarks at non-zero temperature, following Refs.~\cite{Aarts:2005hg,Aarts:2006em}. These results are presented in the Supplementary Material. Here we demonstrate how smearing suppresses both effects.

\begin{figure}[t]
    \centering
    \includegraphics[width=0.96\linewidth]{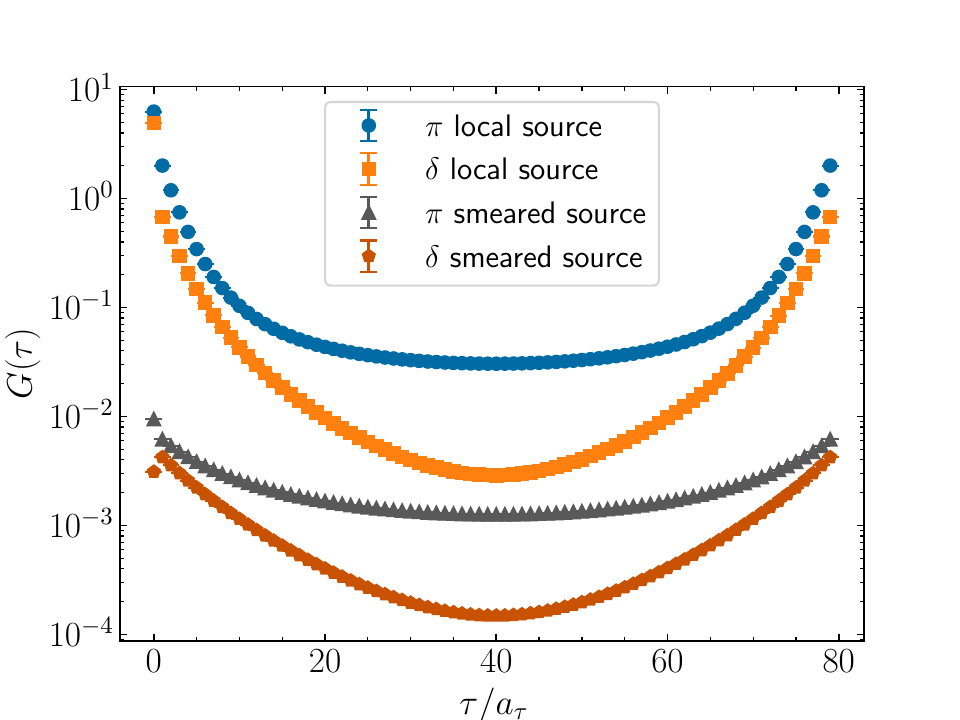}
    \includegraphics[width=0.96\linewidth]{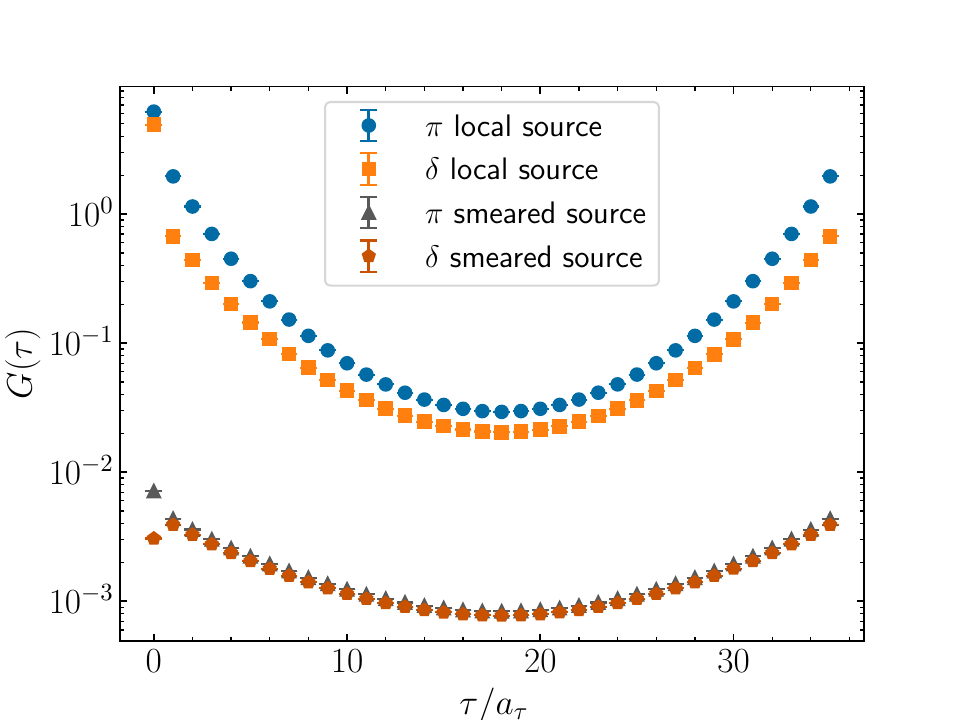}
    \caption{Generation 3 correlators (in lattice units) in the scalar and pseudoscalar channels obtained with local and smeared sources for $T=160$ $\mathrm{MeV}$ (above) and $356$ $\mathrm{MeV}$ (below).}
    \label{fig:correlators}
\end{figure}

\begin{figure}[t]
    \centering
    \includegraphics[width=0.96\linewidth]{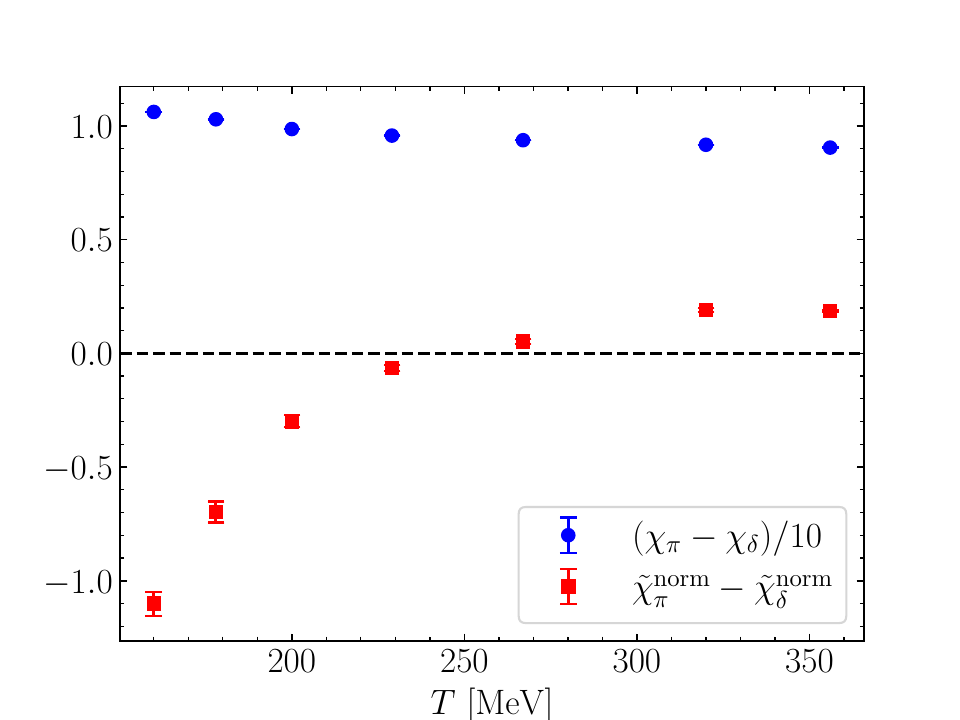}
    \includegraphics[width=0.96\linewidth]{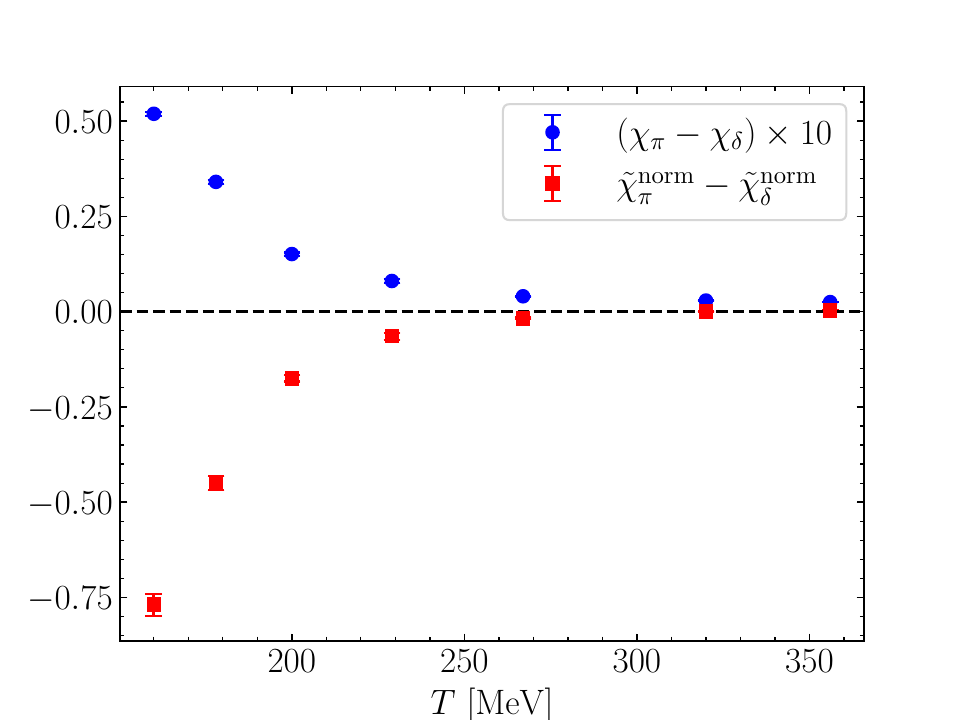}
    \caption{Difference of the susceptibilities $\chi_\pi-\chi_\delta$ and $\widetilde\chi_\pi^{\rm norm}-\widetilde\chi_\delta^{\rm norm}$, see \eqnr{eq:sus} and \eqnr{eq:norm_sus}, for local (above) and smeared (below) correlators on Generation 3. In the upper (lower) plot, $\chi_\pi-\chi_\delta$ is divided (multiplied) by a factor of 10 for visibility. Data show only statistical errors and a single choice of $\tau_{\min}$, $\tau_{\min}T\approx 0.2$}
    \label{fig:susceptibilities}
\end{figure}

In Figure \ref{fig:correlators} we show the scalar and pseudoscalar correlators with local and smeared sources and sinks at a low and a high temperature. The emerging degeneracy is best visible for smeared sources at high temperature and is of course absent at low temperature.

The corresponding differences between the susceptibilities are shown in Figure \ref{fig:susceptibilities}. The difference between the standard susceptibilities, $\chi_\pi-\chi_\delta$, never crosses zero, although results with smeared sources are much closer to zero than with local sources. 
For the normalised difference, $\widetilde\chi_\pi^{\rm norm}-\widetilde\chi_\delta^{\rm norm}$, the one with local sources crosses zero, while with smeared sources it remains zero at the highest temperatures.

\begin{figure}[t]
  \centering
  \includegraphics[width=0.96\linewidth]{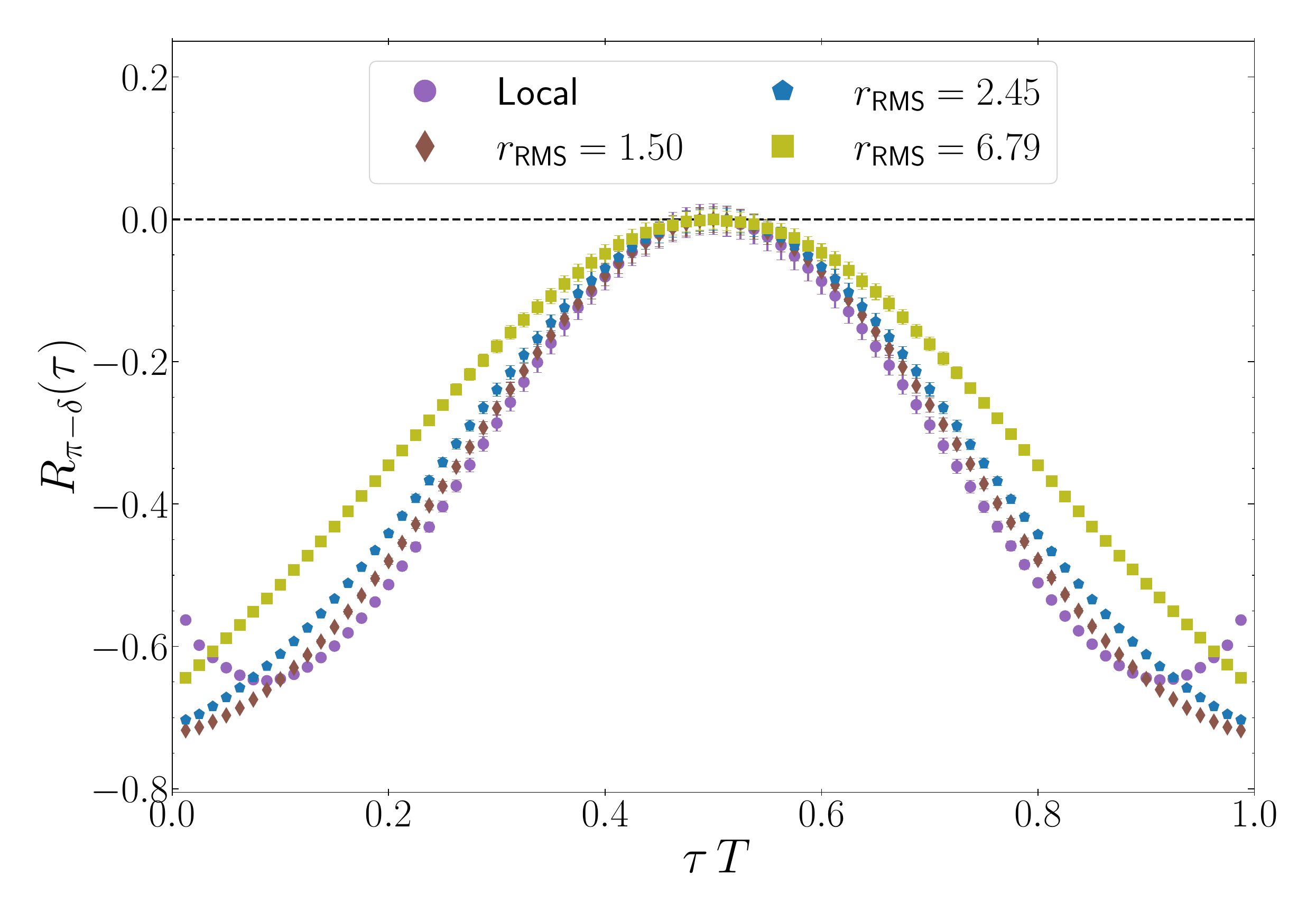} \hfill
  \includegraphics[width=0.96\linewidth]{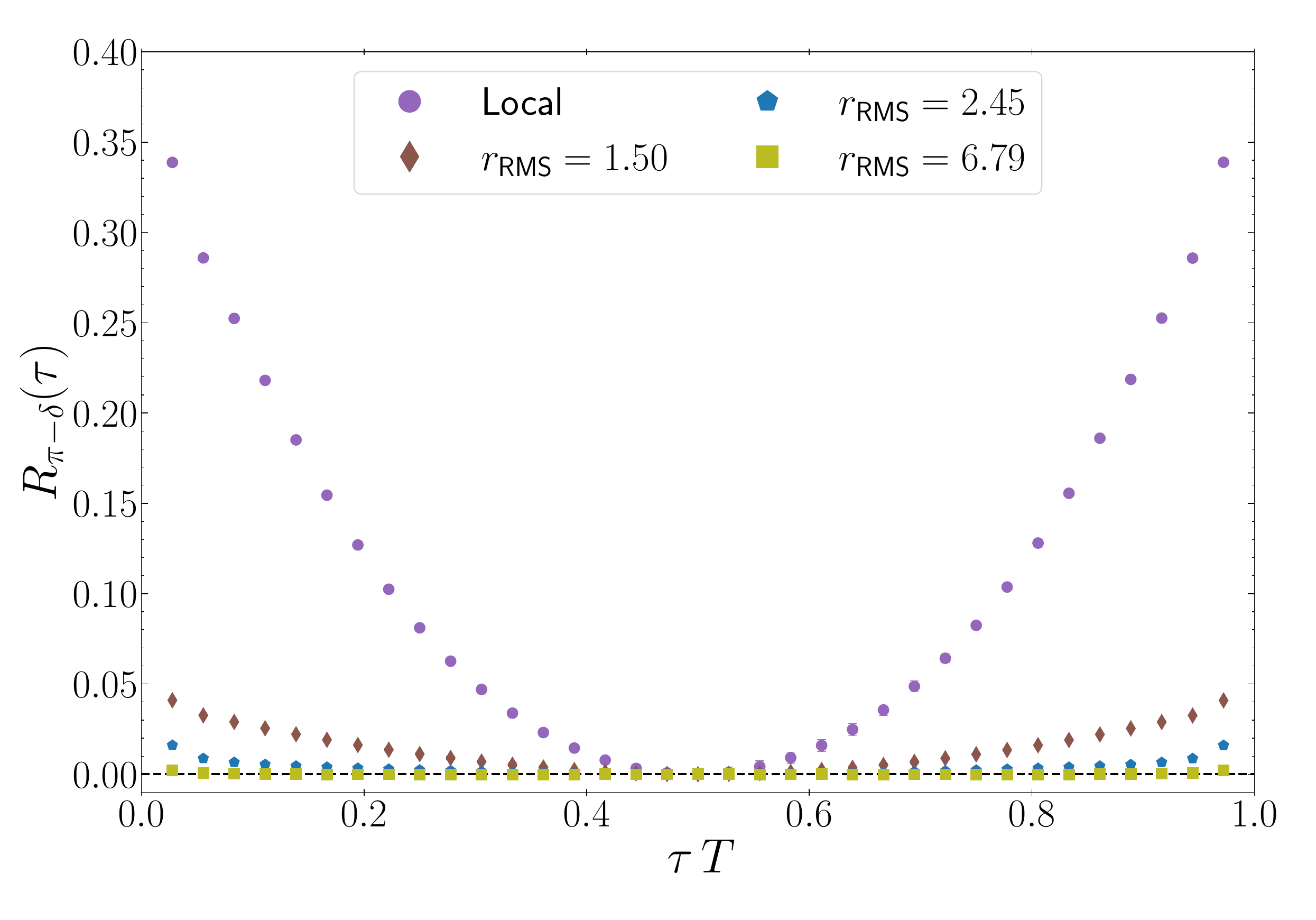}
  \caption{$R_{\pi-\delta}(\tau)$ for four different smearing radii (including none) at $T=160$ MeV (above) and $356$ MeV (below) on Generation 3 ensembles. Note the different vertical scales.}
  \label{fig:smearComp}
\end{figure}

\begin{figure}[b]
    \centering
    \includegraphics[width=0.96\linewidth]{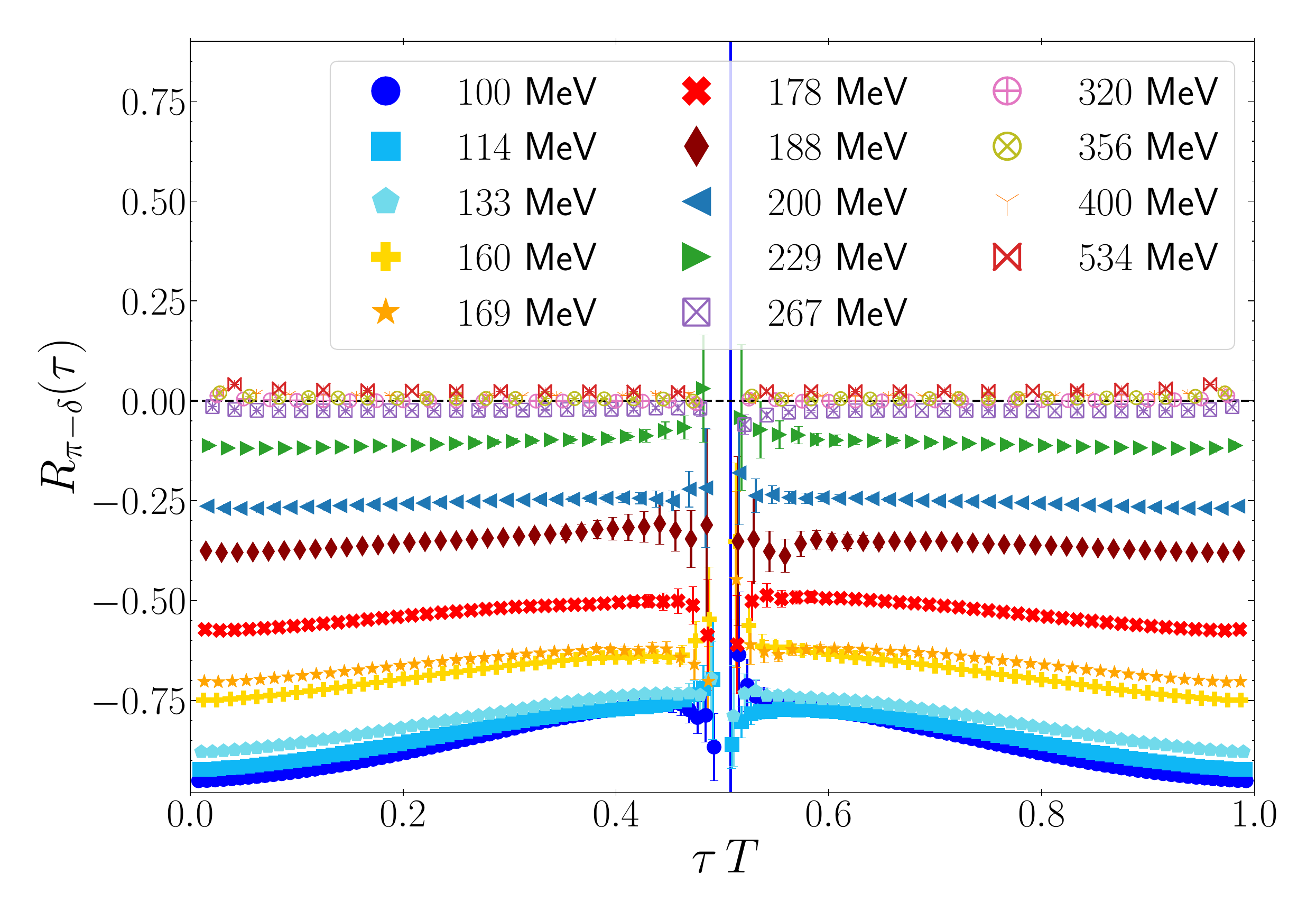}
    \caption{ $R_{\pi-\delta}(\tau)$ using smeared midpoint subtracted correlators for Generation 3. \Fig{fig:PS_R_ratio_smeared} shows the corresponding data without midpoint subtraction.}
    \label{fig:G3_Sub_R}
\end{figure}

To determine the degree of smearing, we study the ratio $R_{\pi-\delta}(\tau)$. Four different smearing levels (including none, i.e., local) are compared in \Fig{fig:smearComp} at two different temperatures. 
The amount of smearing is described by the root-mean-square radius of the Gaussian profile on a free point source. We have previously used the $r_\text{RMS} = 6.79$ profile for nucleon and charm baryon spectroscopy~\cite{Aarts:2020vyb,Aarts:2023nax}. The other profiles were chosen to be smaller.
Smearing makes the ratio flatter at high temperature.
At the lower temperature, the difference between the smeared and local correlators is less evident. Close to the midpoint of the correlator, the ground state should be dominant and so the smeared correlators should be close to the local correlator.  This is evident for smearing radii $r_\text{RMS}=1.50,\,2.45$ but not for $r_\text{RMS}=6.79$. This is not surprising as this larger smearing radius was optimised for nucleon correlation functions~\cite{Aarts:2020vyb}. In our main analysis, we have selected the $r_{\text{RMS}}=2.45$ correlator corresponding to $\kappa=1.96,\,n=6$ in our smearing algorithm~\cite{Gusken:1989qx,Aarts:2023nax}, as it more effectively eliminates the upwards curve at short temporal separation.

\paragraph{Midpoint subtracted correlators --}
At nonzero temperature, a constant contribution may be present in meson correlators, due to single quark propagation around the temporal length of the lattice \cite{Umeda:2007hy,Ohno:2011zc,Arikawa:2025kjx,Urrea-Nino:2025ijv}.
This can be resolved by subtracting the midpoint of the correlator, i.e., by considering $\tilde{G}(\tau) = G(\tau) - G\left(N_\tau / 2\right)$. 
To check the robustness of our results, we repeat the analysis of \Fig{fig:PS_R_ratio_smeared} using the midpoint subtracted correlators.
Both the pseudoscalar $\left(\pi\right)$ and scalar $\left(\delta\right)$ correlators are normalised at their midpoint before the midpoint is subtracted.
The result is shown in \Fig{fig:G3_Sub_R}.
We observe that the ratio approaches zero, as before, at temperatures above 267 MeV, confirming the robustness of our analysis.

\clearpage

\section{Supplementary Material}
\label{sec:supp}

\paragraph{Chiral symmetry restoration --}
In our previous work, the transition temperature for chiral symmetry restoration, denoted as $T_c$, was determined via the inflection point of the renormalised chiral condensate, 
computed using local operator sources and without a temporal cut, see Refs.~\cite{Aarts:2014nba,aarts_2023_8403827,Aarts:2020vyb,aarts_2024_10636046}. This leads to the values of $T_c$ quoted in Table \ref{tab:lattice}. Here we demonstrate that the transition temperature can also be determined using a similar analysis as used in the main paper for effective $U(1)_A$ restoration, albeit with a larger uncertainty.

\begin{figure}
    \centering
    \includegraphics[width=1.0\linewidth]{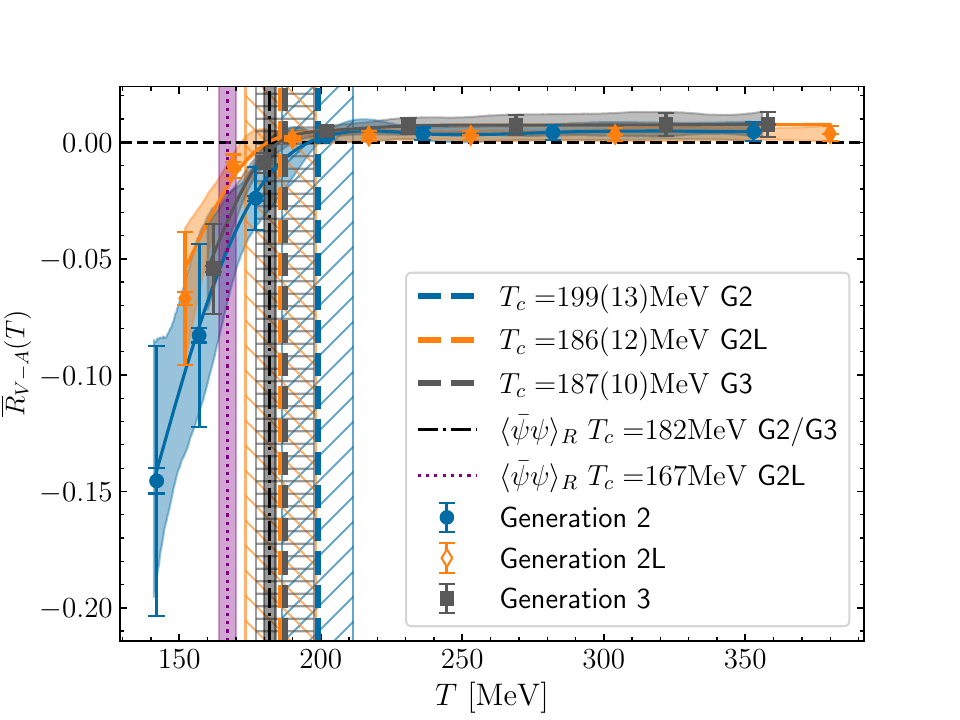}
    \caption{$\overline{R}_{V-A}$ results for Generation 2 $(\text{blue circles }\textcolor{tab10Blue}{\bigcirc})$, 2L $(\text{orange diamonds }\textcolor{tab10Orange}{\diamond})$ and 3 $(\text{dark-gray squares }\textcolor{tab10Gray}{\square})$. The filled curves show cubic spline interpolations for each dataset; for each generation, the transition temperature $T_{c}$ is determined as the temperatures where the spline reaches zero.}
    \label{fig:supp:chiral_R}
\end{figure}

Chiral symmetry relates the vector ($\rho$) and axial-vector ($a_1$) mesons. We compute correlators as in Eq.~\eqref{eq:sus-default}, with $\Gamma=\gamma_i$ and $\gamma_5\gamma_i$ respectively, and construct the normalised susceptibilities using smeared correlators as before. Since we take the difference between vector and axial-vector channels, we denote the ratio as $\overline{R}_{V-A}$. The results are shown in Fig.~\ref{fig:supp:chiral_R}, for all three generations. The transition temperatures are somewhat higher and the uncertainties are about 10 MeV larger than those obtained using the renormalised chiral condensate. The results for Generation 3 are compatible within errors; the ability to carry out a finer scan over the temperatures is essential in this approach. We observe that $\overline{R}_{V-A}$ makes a rather sharp turnover after crossing $T_c$. When the temperature scan is not fine enough, this quick turnover may not be picked up by the spline, leading to a shift in $T_c$ to a higher value.
We have also carried out the analysis using local sources and found that in that case the results are closer to the ones obtained using the renormalised chiral condensate~\cite{Smecca:2024gpu}, while they suffer from larger systematic errors due to the choice of $\tau_{\min}$. In any case, the chiral transition temperature and the temperature for effective $U(1)_A$ restoration are clearly separated, irrespective of the method used to determine them.

\paragraph{Correlators in the non-interacting case --}
Additional insight into the effect of smearing and the Wilson mass term can be gained by comparing with results obtained from non-interacting Wilson fermions, on lattices with the same geometry as the Generation 3 ensembles (similar results are obtained for Generation 2 and 2L), following Ref.~\cite{Aarts:2005hg}. We also make a comparison with non-interacting overlap fermions, following Ref.~\cite{Aarts:2006em}.

Comparing the results for $R(\tau)$ using smeared and local sources and sinks, see 
Figs.~\ref{fig:PS_R_ratio_smeared} and \ref{fig:PS_R_ratio_local}, we observed that in the smeared case the upward curvature at the early times, say $\tau T\lesssim 0.2$, is suppressed, compared to the local case. Where this curvature sets in influences the choice of $\tau_{\min}$ and affects systematic uncertainties. 

\begin{figure}
    \centering
    \includegraphics[width=0.95\linewidth]{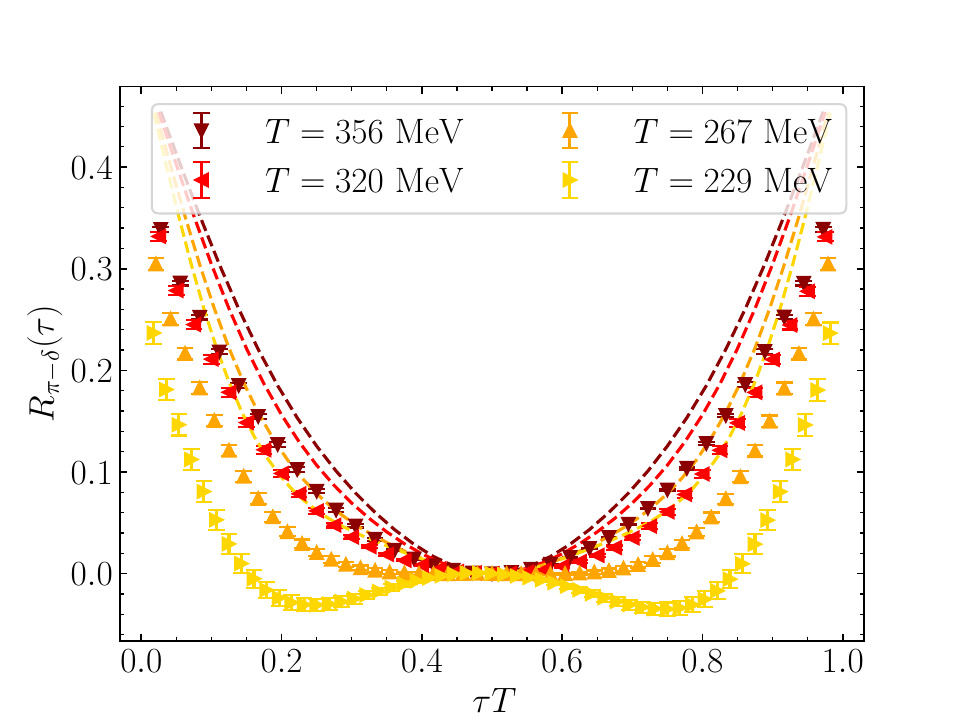}
    \includegraphics[width=0.95\linewidth]{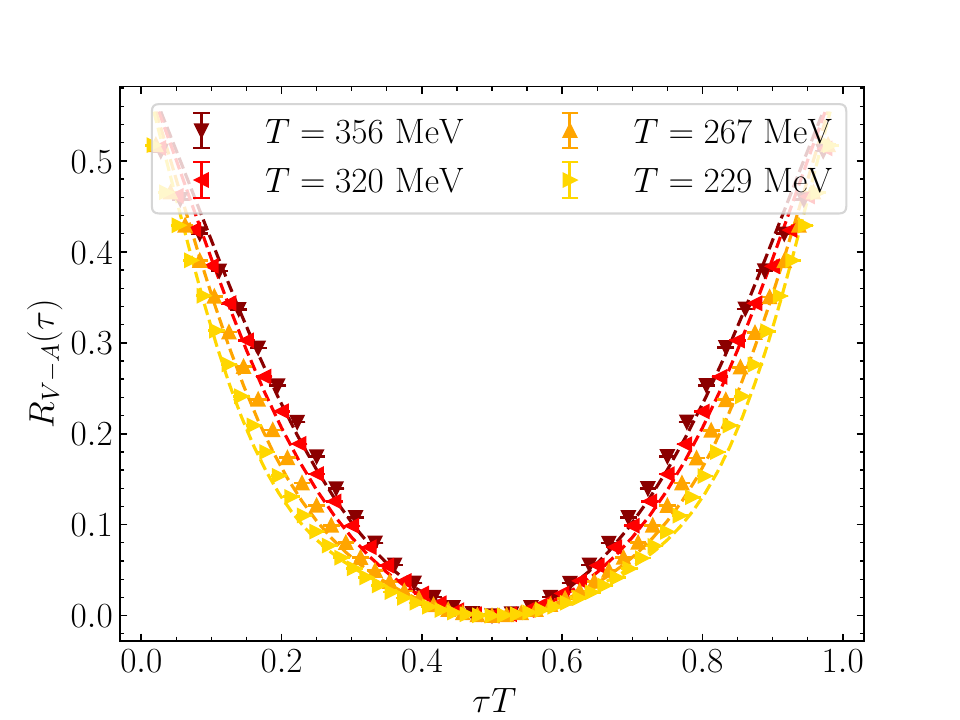}
    \caption{$R_{\pi-\delta}(\tau)$ (above) and $R_{V-A}(\tau)$ (below) at high temperature using local correlators for Generation 3 (symbols) and using non-interacting lattice fermions (dashed lines). 
    }
    \label{fig:supp:R_free}

    \centering
    \includegraphics[width=0.95\linewidth]{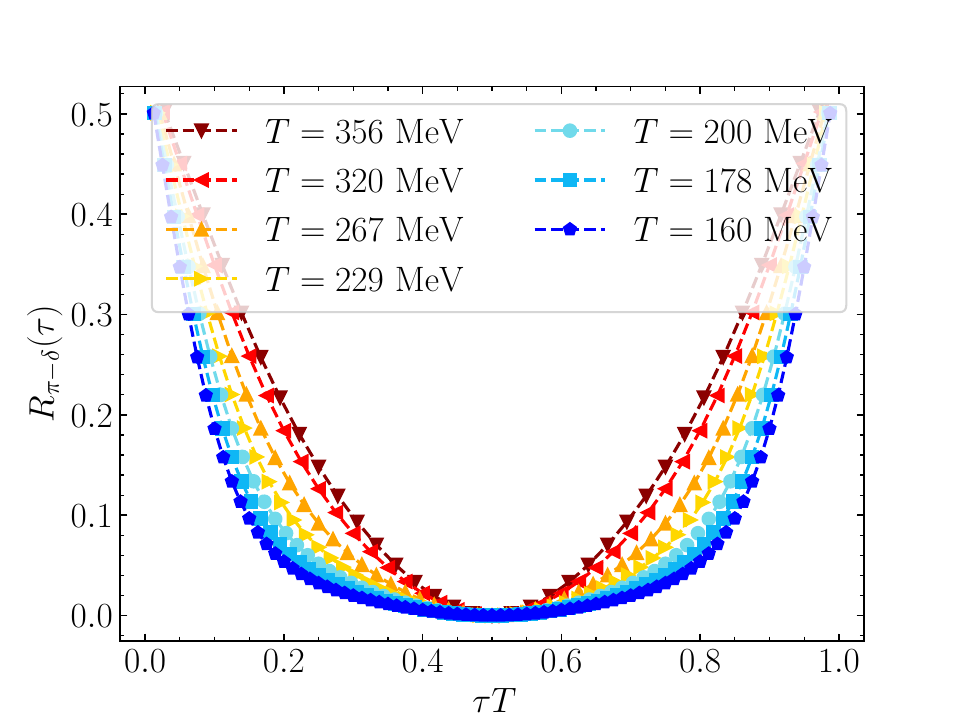}
    \caption{$R_{\pi-\delta}(\tau)$ for non-interacting massless Wilson fermions, using the Generation 3 geometry.}
    \label{fig:supp:free_massless_Wilson_Overlap}
\end{figure}

We now demonstrate that the upward curvature has its origin in the Wilson mass term and hence reflects the explicit breaking of chiral symmetry at nonvanishing lattice spacing. To show this, we compute $R(\tau)$ using non-interacting Wilson fermions on the same lattices as in Generation 3, using the same bare mass parameter and anisotropy. 
In \Fig{fig:supp:R_free} we show $R_{\pi-\delta}(\tau)$ (scalar-pseudoscalar) and $R_{V-A}(\tau)$ (vector--axial-vector) at the higher temperatures. Data points are Generation 3 results using local correlators; dashed lines indicate the non-interacting results. 
We observe that the upward curvature is also present in the free case, confirming its origin.
As a side comment, in the vector--axial-vector case the QCD ratios are much closer to the non-interacting values within this temperature range, indicating more remnants of the strong interaction in the scalar-pseudoscalar channels compared to the vector-axial--vector channels. 
For a comparison between screening masses in the pseudoscalar and vector channels at very high temperature, see e.g.\ Ref.~\cite{DallaBrida:2021ddx}.

In \Fig{fig:supp:free_massless_Wilson_Overlap} we show the same ratio $R_{\pi-\delta}(\tau)$ for non-interacting {\em massless} Wilson fermions, indicating that the curvature is also present in this case. We have verified that for non-interacting massless overlap fermions, with exact chiral symmetry at nonzero lattice spacing, $R_{\pi-\delta}(\tau)$ is strictly zero, as it should be. 

\begin{figure}
    \centering
    \includegraphics[width=0.95\linewidth]{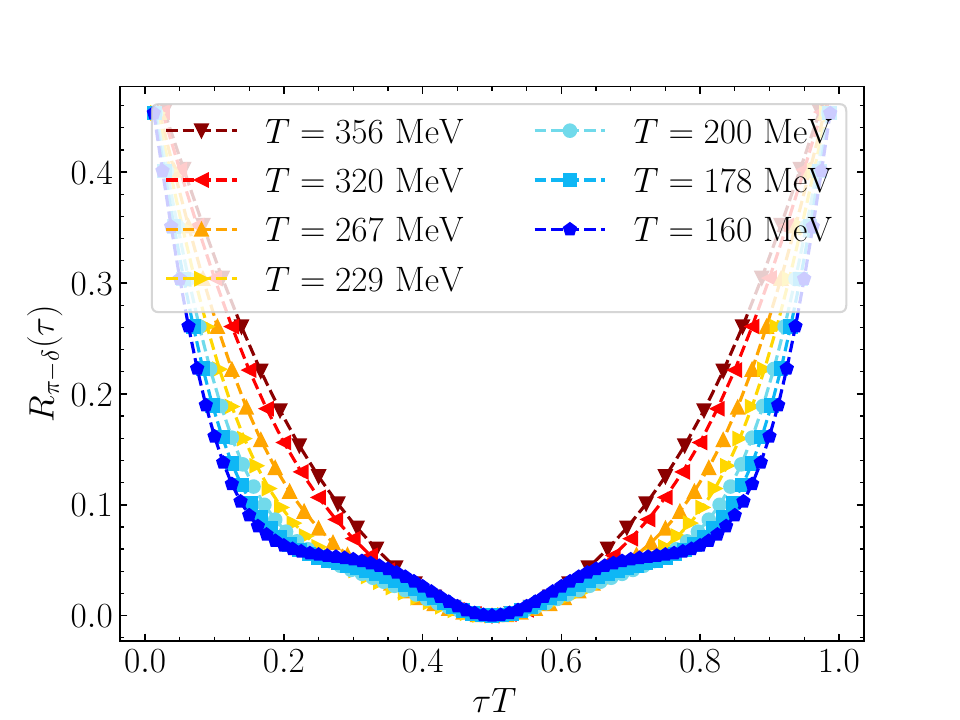}
    \includegraphics[width=0.95\linewidth]{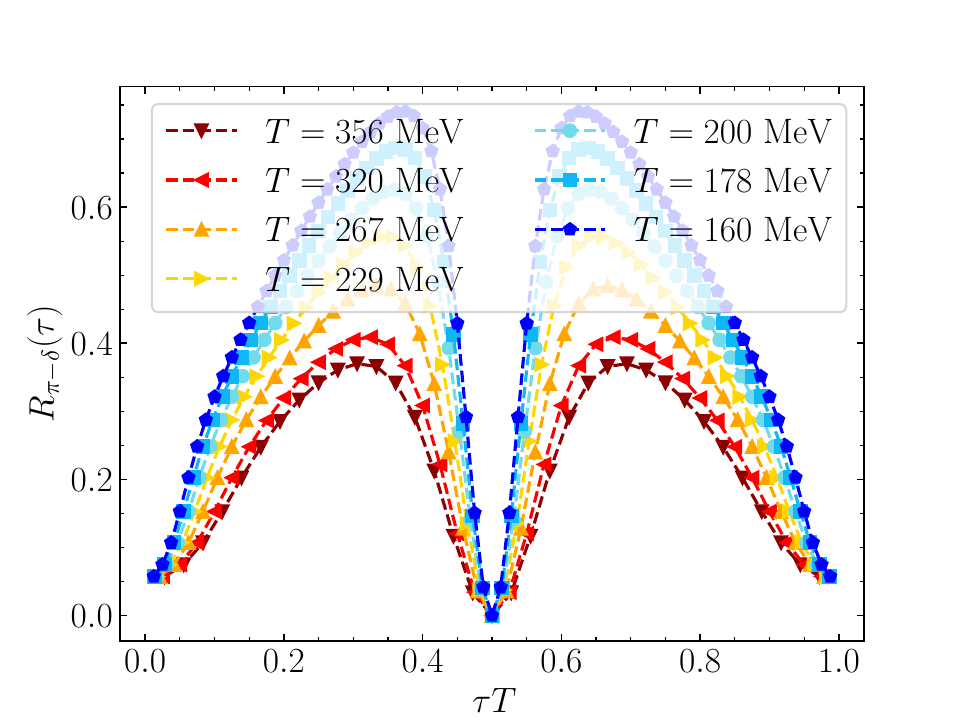}
    \caption{$R_{\pi-\delta}(\tau)$ for non-interacting massive Wilson (above) and overlap (below) fermions, using Generation 3 geometry.}
    \label{fig:supp:free_Wilson_Overlap}
\end{figure}

Finally, in Figure~\ref{fig:supp:free_Wilson_Overlap} we show the same ratio for massive Wilson and overlap fermions, across the entire temperature range. Note that for Wilson fermions the data points at the four highest temperatures correspond to the dashed lines in \Fig{fig:supp:R_free}. Here we observe a non-convex ratio for Wilson fermions at the lower temperatures, indicating an interplay between the bare mass and the Wilson term, and a deviation from zero for overlap fermions, as expected. We note that these results for overlap fermions are obtained using parameters which do not resemble the Wilson fermion results, and therefore the magnitude of $R(\tau)$ should be considered only a qualitative rather than a quantitative estimate of chiral symmetry breaking.

\bibliography{main}
\bibliographystyle{JHEP}

\end{document}